\documentclass[11pt]{article}
\usepackage{amssymb, amsfonts, latexsym, amsthm, amsmath}
\usepackage[left=1.8cm,right=1.8cm,top=1.6cm,nohead]{geometry}
\pdfoutput=1

\newcommand{\Ndist}{\mathcal{N}}

\newcommand{\abs}[1]{\left| #1 \right|}

\newcommand{\deriv}[2]{\frac{d#1}{d#2}}

\newcommand{\half}{\frac{1}{2}}

\newcommand{\vc}[1]{\mathbf{#1}}

\newcommand{\dsqom}{\frac{1}{\sqrt{\Omega }}}

\begin{document}
\title{A simplified derivation of the Linear Noise Approximation}
\author{Edward W. J. Wallace}
\date{\today}
\maketitle

\begin{abstract}
Given a discrete stochastic process, for example a chemical reaction system or a birth and death process, we often want to find a continuous stochastic approximation so that the techniques of stochastic differential equations may be brought to bear. One powerful and useful way to do this is the system size expansion of van Kampen to express a trajectory as a small stochastic perturbation to a deterministic trajectory, using a small parameter related to the volume of the system in question. This is usually pursued only up to first order, called the Linear Noise Approximation. The usual derivation of this proceeds via the master equation of the discrete process and derives a Fokker-Planck equation for the stochastic perturbation, both of which are equations for the evolution of probability distributions. Here we present a derivation using stochastic difference equations for the discrete process and leading, via the chemical Langevin equation of Gillespie, directly to a stochastic differential equation for the stochastic perturbation. The new derivation, which does not yield the full system size expansion, draws more explicitly on the intuition of ordinary differential equations so may be more easily digestible for some audiences.
\end{abstract}

\section{Introduction}

Given a discrete stochastic process, for example a chemical reaction system or a birth and death process, we often want to find a continuous stochastic approximation so that the techniques of stochastic differential equations may be brought to bear. One powerful and useful way to do this is the system size expansion of Van Kampen \cite{van-kampen2007}, which has been proved useful in applications in biophysics \cite{elf2003}, epidemiology \cite{alonso2007}, ecology \cite{mckane2005} and neuroscience \cite{benayoun2010,bressloff2009}.
 If the discrete process has a size or volume parameter $\Omega$ for which reaction rates scale appropriately, $1/\sqrt{\Omega}$ may be used as a small parameter. Inspired by the central limit theorem, we approximate the discrete random variable $K$ as the sum of a continuous deterministic variable $x$, scaled with $\Omega$, and a smaller continuous stochastic perturbation $\xi$ scaled with $\sqrt{\Omega}$, in other words $K = \Omega x + \sqrt{\Omega} \xi$. The chief application for this is to derive a linear noise approximation, meaning that the expansion is pursued to first order in $1/\sqrt{\Omega}$, in which the stochastic perturbation satisfies a Fokker-Planck equation with linear drift term and constant noise term, both calculated from the deterministic solution. The small parameter may be used to check the quality of the approximation, calculate higher-order corrections to moments, and so on, but this is almost never done.

The usual method of deriving the system size expansion is
\begin{enumerate}
  \item Express the discrete process in terms of its master equation.
  \item Make the ansatz $K = \Omega x + \sqrt{\Omega} \xi$.
  \item Re-express the difference terms in the master equation in terms of differential operators in the new variables.
  \item Taylor expand the coefficients of the resulting equation about the deterministic term $x$.
  \item Collect terms to find a deterministic $O(\sqrt{\Omega})$ equation for $x$ and a Fokker-Planck equation for $\xi$ with linear terms of $O(1)$, and terms of higher order in $1/\sqrt{\Omega}$.
\end{enumerate}
While this is a perfectly rigorous way to derive the approximation, it involves a lot of notation, and can be a challenge to present to audiences who are unfamiliar with the master equation or Fokker-Planck equation. Many scientists and mathematicians are familiar with ordinary differential equations (ODEs), and with the idea of a stochastic differential equation (SDE) as an ODE with a random input, but are unused to thinking of evolution equations for probability distributions, and the extra steps involved in explaining such evolution equations may prove an obstacle to  the success of a talk or exposition.

Here I present an alternative derivation of the linear noise approximation phrased entirely in SDE notation. Given the same scaling hypothesis regarding transition rates, we may make use of the chemical Langevin equation, or $\tau$-leaping, due to Gillespie~\cite{gillespie2001,gillespie2007}.  This relies on a timescale $\tau$ at which there are many transitions per channel but over which the reaction rates do not change appreciably, so that the number of reactions per channel is approximately Gaussian over that timescale. The derivation proceeds
\begin{enumerate}
  \item Express the discrete process in terms of its transition rates.
  \item Given an appropriate timescale $\tau$,  re-express the increment random variables as Gaussians, giving rise to an SDE.
  \item Make the ansatz $K = \Omega x + \sqrt{\Omega} \xi$.
  \item Taylor expand the coefficients of the resulting equation about the deterministic term $x$.
  \item Collect terms to find a deterministic $O(\sqrt{\Omega})$ equation for $x$ and a Langevin equation for $\xi$, with linear terms of $O(1)$, and terms of higher order in $1/\sqrt{\Omega}$.
\end{enumerate}
The deterministic variables $x$ we arrive at satisfy the same equations and so are identical. Using the \^Ito interpretation, which is reasonable for systems with internal noise~\cite{van-kampen2007}, the stochastic perturbation $\xi$ also corresponds in the two derivations to linear order.

\section{Derivation for chemical reaction systems}

Let $\vc{K}(t)$ be a vector of the numbers of different chemical species, subject to reaction channels $\vc{K} \longrightarrow \vc{K} + \vc{v}_j$ each at rate ${\Omega r_j(\vc{K}/\Omega)}$, where the $r_j$'s are polynomials or other suitably nice functions. Note that we have built the system-size parameter into the reaction rates; discussions of why this is reasonable may be found in \cite{gardiner2009,van-kampen2007}. Now suppose there is a timescale $\tau$ at which the number of reactions per channel is large but the reaction rates $r_j$ do not change appreciably. Known as $\tau$-leaping, this is discussed extensively in \cite{gillespie2007, gillespie2009}; under our scaling hypothesis this occurs roughly when 
$\displaystyle{ \frac{1}{\Omega r_j} \ll \tau \ll \abs{ \frac{ r_j \tau}{ \Delta_\tau r_j } }
}$, where $\Delta_\tau$ indicates the change in a quantity in time $\tau$. Since the number of reactions per channel is then approximately normal and independent of the number in other channels, the evolution of $\vc{K}(t)$ is described by
\begin{align}
\vc{K}(t+ \tau) 
& \approx \vc{K}(t) + \sum_j  \Ndist \left( \tau \Omega r_j(\vc{K}/\Omega), \tau \Omega r_j(\vc{K}/\Omega) \right) \vc{v}_j \\
& =  \vc{K}(t) +  \tau \Omega \sum_j r_j(\vc{K}/\Omega) \vc{v}_j
+ \sqrt{ \tau \Omega} \sum_j \sqrt{r_j(\vc{K}/\Omega)} \, \Ndist \left(0,1 \right) \vc{v}_j \label{eq:Ktauevol}
\end{align}
which is the ``chemical Langevin equation'' described in \cite{gillespie2000}.

Now we make the ansatz of $\vc{K} = \Omega \vc{x} + \sqrt{\Omega} \vc{\xi}$, a deterministic solution with a stochastic perturbation. Then \eqref{eq:Ktauevol}, divided through by $\Omega$, becomes
\begin{multline}
\label{eq:fulltauevol}
\vc{x}(t+ \tau) + \dsqom \vc{\xi}(t+\tau)
\\ =  \vc{x}(t) + \dsqom \vc{\xi} +  \tau \sum_j r_j \left(\vc{x}(t) + \dsqom \vc{\xi} \right)  \vc{v}_j
+ \sqrt{\frac{ \tau }{\Omega}} \sum_j \sqrt{r_j \left(\vc{x}(t) + \dsqom\vc{\xi} \right)} \, \Ndist \left(0,1 \right) \vc{v}_j \text{ .}
\end{multline}
As promised, we now Taylor expand the reaction rates so that
\begin{equation}
\label{eq:TaylorRates}
r_j \left( \vc{x} + \dsqom \vc{\xi} \right)
= r_j(\vc{x} ) +  \dsqom   D_\vc{x} r_j  (\vc{\xi}) 
+ \frac{1}{\Omega }  \half D_\vc{x}^2 r_j  (\vc{\xi}) + \ldots
\end{equation}
where we denote by $D_\vc{x}^m r_j  (\vc{\xi})$ the $m$th directional derivative of $r_j$, evaluated at $\vc{x}$, in the direction of $\vc{\xi}$. Substituting this in \eqref{eq:fulltauevol}, we obtain 
\begin{multline}
\label{eq:taylortauevol}
\vc{x}(t+ \tau) + \frac{1}{\sqrt{\Omega }} \vc{\xi}(t+\tau)
\\ =  \vc{x}(t) + \frac{1}{\sqrt{\Omega }}\vc{\xi}(t) +  \tau \sum_j \left[ r_j(\vc{x} ) +  \frac{1}{\sqrt{\Omega} }   D_\vc{x} r_j  (\vc{\xi}(t)) 
+ \frac{1}{\Omega }  \half D_\vc{x}^2 r_j  (\vc{\xi}(t)) + \ldots \right] \vc{v}_j \\
+ \sqrt{\frac{ \tau }{\Omega}} \sum_j \sqrt{\left[ r_j(\vc{x} ) +  \frac{1}{\sqrt{\Omega} }   D_\vc{x} r_j  (\vc{\xi}(t)) 
+ \frac{1}{\Omega }  \half D_\vc{x}^2 r_j  (\vc{\xi}(t)) + \ldots \right]} \, \Ndist \left(0,1 \right) \vc{v}_j \text{ .}
\end{multline}
Collecting terms of $O(1)$ in \eqref{eq:taylortauevol}, we have a deterministic expression for the concentration term $x$
\begin{equation}
\label{eq:xtauevol}
\vc{x}(t+ \tau) = \vc{x}(t) + \tau \sum_j  r_j(\vc{x} ) \vc{v}_j
\end{equation}
which may be expressed as an ordinary differential equation
\begin{equation}
\label{eq:xevol}
 \deriv{ \vc{x}(t)}{t} =\sum_j  r_j(\vc{x} ) \vc{v}_j
\end{equation}
which is the reaction rate equation.
Collecting terms of $O\left(\dsqom\right)$ and higher, we obtain a stochastic expression for the perturbation term $\xi$ whose coefficients are evaluated at the deterministic solution $\vc{x}(t)$,
\begin{multline}
\label{eq:xitauevol}
  \vc{\xi}(t+\tau)
=   \vc{\xi}(t) +  \tau \sum_j \left[   D_\vc{x} r_j  (\vc{\xi}(t)) 
+   \dsqom  \half D_\vc{x}^2 r_j  (\vc{\xi}(t)) + \ldots \right] \vc{v}_j \\
\\
+ \sqrt{ \tau } \sum_j \sqrt{\left[ r_j(\vc{x} ) +  \dsqom  D_\vc{x} r_j  (\vc{\xi}(t)) 
+ \frac{1}{\Omega }  \half D_\vc{x}^2 r_j  (\vc{\xi}(t)) + \ldots \right]} \, \Ndist \left(0,1 \right) \vc{v}_j 
\text{ .}
\end{multline}
We may also write this equation in Langevin form, using the \^Ito interpretation
\begin{multline}
\label{eq:xievol}
  \deriv{\vc{\xi}(t)}{t}
=   \sum_j \left[  D_\vc{x} r_j  (\vc{\xi}(t)) 
+   \dsqom   \half D_\vc{x}^2 r_j  (\vc{\xi}(t)) + \ldots \right] \vc{v}_j \\
\\
+ \sum_j \sqrt{\left[ r_j(\vc{x} ) +  \dsqom  D_\vc{x} r_j  (\vc{\xi}(t)) 
+ \frac{1}{\Omega }  \half D_\vc{x}^2 r_j  (\vc{\xi}(t)) + \ldots \right]} \, \eta_j \vc{v}_j 
\end{multline}
where the $\eta_j$ are independent white-noise variables, one for each reaction channel. It would be possible to replace these with one independent white noise per chemical species, calculating the coefficients from the last term in \eqref{eq:xievol}.

Neglecting the terms in $O \left( \dsqom \right)$ and higher, we reproduce the linear noise approximation
\begin{equation}
\label{eq:xiLNA}
  \deriv{\vc{\xi}(t)}{t}
=   \sum_j  D_\vc{x} r_j  (\vc{\xi}(t))  \vc{v}_j 
+ \sum_j \sqrt{ r_j(\vc{x} ) } \, \eta_j \vc{v}_j 
\text{ .}
\end{equation}
This detailed derivation might still present a challenge to explain to a mixed audience; since the system size expansion is almost always truncated at the linear noise approximation, one could omit terms of order $\frac{1}{\Omega}$ and higher earlier on, and write down the linear noise approximation directly from \eqref{eq:fulltauevol}.

The derivation presented here comes from the chemical Langevin equation, which is formally equivalent to the Fokker-Planck equation, and so could not be used to make the full system size expansion, which is derived from the infinite-order Kramers-Moyal equation.

\section{Derivation for birth-death processes}

The derivation simplifies slightly for birth-death processes, i.e. discrete processes in which the only transitions are $\vc{K} \longrightarrow \vc{K} \pm \vc{e}_i$, where $\vc{e}_i$ is the vector incrementing the $i$th population by 1. These can also be thought of as (inhomogenous) random walks on a square lattice. Our scaling assumption is incorporated by specifying the rate of the transition $\vc{K} \longrightarrow \vc{K} \pm \vc{e}_i$ as $\Omega t^\pm_i(\vc{k}/\Omega)$.

Again suppose there is a timescale $\tau$ at which the number of reactions per channel is large but the size-scaled reaction rates $t_j$ do not change appreciably; this occurs roughly when 
$\displaystyle{ \frac{1}{\Omega t_j} \ll \tau \ll \abs{ \frac{ t_j \tau}{ \Delta_\tau t_j } }
}$, where $\Delta_\tau$ indicates the change in a quantity in time $\tau$.
 In this case the increment random variables
 are approximately normal, and the evolution of $\vc{K}(t)$ is described by
\begin{align}
\label{eq:bdKtauevol}
\vc{K}(t+ \tau) 
& \approx \vc{K}(t) + \sum_i  \Ndist \left( \tau \Omega t^+_i \vc{K}/\Omega), \tau \Omega t^+_i (\vc{K}/\Omega) \right) \vc{e}_i 
 - \sum_i  \Ndist \left( \tau \Omega t^-_i \vc{K}/\Omega), \tau \Omega t^-_i (\vc{K}/\Omega) \right) \vc{e}_i
 \\
& =  \vc{K}(t) +  \tau \Omega \sum_i \left[ t^+_i(\vc{K}/\Omega) - t^-_i(\vc{K}/\Omega)  \right] \vc{e}_i
+ \sqrt{ \tau \Omega} \sum_i \sqrt{ \left[ t^+_i(\vc{K}/\Omega) + t^-_i(\vc{K}/\Omega)  \right]  } \, \Ndist \left(0,1 \right) \vc{e}_i 
\\
& = \vc{K}(t) +  \tau \Omega \sum_i A_i(\vc{K}/\Omega) \vc{e}_i
.
+ \sqrt{ \tau \Omega} \sum_i \sqrt{B_i(\vc{K}/\Omega) } \, \Ndist \left(0,1 \right) \vc{e}_i 
\end{align}
Since the Gaussian increments for the increase and decrease of each population are independent, we have combined them into a single Gaussian for each population, and introduced the drift and diffusion vectors respectively,
\begin{align}
A_i(\vc{x}) & = t^+_i(\vc{x} ) - t^-_i(\vc{x} ), \\
 B_i(\vc{x}) & = t^+_i(\vc{x} ) + t^-_i(\vc{x} ) .
\end{align}
Making the ansatz $\vc{K} = \Omega \vc{x} + \sqrt{\Omega} \vc{\xi}$, we obtain equations
\begin{align}
\label{eq:bdxtauevol}
\vc{x}(t+ \tau) & = \vc{x}(t) + \vc{A(x)} \tau
\end{align}
which translates to the ordinary differential equation
\begin{equation}
\label{eq:bdxevol}
 \deriv{ \vc{x}(t)}{t} = \vc{A(x)} \text{ .}
\end{equation}
The stochastic perturbation $\xi$ evolves as
\begin{multline}
\label{eq:bdxitauevol}
  \vc{\xi}(t+\tau)
=   \vc{\xi}(t) +  \tau \sum_i \left[   \left( D_\vc{x} A_i \right)(\vc{\xi}(t)) 
+   \dsqom  \half \left(D_\vc{x}^2 A_i \right) (\vc{\xi}(t)) + \ldots \right] \vc{e}_i \\
\\
+ \sqrt{ \tau } \sum_i \sqrt{\left[ B_i 
+  \dsqom \left( D_\vc{x}  B_i \right) (\vc{\xi}(t)) 
+ \frac{1}{\Omega }  \half \left( D_\vc{x}^2  B_i \right) (\vc{\xi}(t)) + \ldots \right]} \, \Ndist \left(0,1 \right) \vc{e}_i
\end{multline}
where the coefficients are evaluated at the deterministic solution of \eqref{eq:bdxevol}, $A_i = A_i(\vc{x}(t))$ and $B_i = A_i(\vc{x}(t))$.
We may again write this equation in Langevin form, using the \^Ito interpretation. Let us pass directly to the Langevin form of the linear noise approximation, neglecting terms of order $\dsqom$, and introducing the Jacobian
\begin{equation}
\label{eq:bdjacobian}
J_{ij} = \partial_i A_j(\vc{x}(t)) =\partial_i \left[  t^+_j(\vc{x}(t) ) - t^-_j(\vc{x} (t)) \right] .
\end{equation}
The linear noise approximation is then
\begin{equation}
\label{eq:bdLNA}
  \deriv{\vc{\xi}(t)}{t}
=   \vc{J} \vc{\xi}(t) + \sum_i \sqrt{B_i} \eta_i(t)
\end{equation}
where $\eta_i(t)$ is a vector of independent white noise variables, one for each population. 
Amongst the strengths of this derivation is to show clearly how the lack of correlations between the white noise inputs to the linear noise approximation \eqref{eq:bdLNA} arises from the elementary transitions in a birth-death process incrementing only one population at a time.

Another simplification is, if the deterministic solution \eqref{eq:bdxevol} is at a fixed point, the drift term $\vc{A} = \vc{0}$; since $A_i(\vc{x})  = t^+_i(\vc{x} ) - t^-_i(\vc{x} ) = 0$, we have $B_i(\vc{x})  = 2t^+_i(\vc{x} ) = 2t^-_i(\vc{x} )$. This means one may choose how to express the noise amplitude $B_i$ at a fixed point, according to whether $t^+_i$ or $t^-_i$ has a simpler form.

\section{Discussion}

This note is intended to clarify a technical point in deriving a tractable continuous approximation for discrete stochastic processes. The linear noise approximation is more precise than the general weak noise approximation described in~\cite{gillespie1992} since the system size supplies a physically-motivated small parameter. The derivation of the linear noise approximation presented here suggests also that the conditions for the chemical Langevin equation to hold are generally also needed for the linear noise approximation to hold. In particular, this requires a "physical" condition relating a timescale to the system size and transition rates. This is in contrast to some of the bolder promises made about the system size expansion: since the system size expansion goes to infinitely many orders in a small parameter, it appears that one may improve its accuracy by proceeding to higher orders in the expansion. However, this is never done; the author is aware of very few pieces of work that make use of higher orders in the system-size or Kramers-Moyal expansions~\cite{risken1987}, and this work does not deal with an application. Given that uses of the system size expansion invariably stop at the linear noise approximation, one may as well derive it as directly and comprehensibly as possible.

\subsection*{Acknowledgments}

The author is grateful to Marc Benayoun and Dan Gillespie for detailed discussions that corrected some errors and improved this note a great deal.

\bibliographystyle{plain} 
\bibliography{/Users/edwardwallace/Documents/Work/2010}

\end{document}